\documentclass{elsart}

\usepackage{graphicx}
\usepackage{float}
\usepackage{amsmath}

\begin{document}
\renewcommand\floatpagefraction{.85} \renewcommand\topfraction{.85}
\renewcommand\bottomfraction{.85} \renewcommand\textfraction{.15}


\newcommand\orbit[1]{\langle #1 \rangle}

\newcommand\perpoint[1]{\overline{#1}}

\newcommand\sym[1]{\text{``}#1\text{''}}

\newcommand\name{\mathcal{N}}
\newcommand\nameof[1]{\name(#1)}
\newcommand\nameofi[2]{\name_{#2}(#1)}

\newcommand\pname{N}
\newcommand\pnameof[2]{N_{#2}(#1)}
\newcommand\pa{\Gamma}

\newcommand\region{\Delta}

\newcommand\border{\beta}
\newcommand\borderof[1]{\border(#1)}
\newcommand\borderofi[2]{\border_{#2}(#1)}

\newcommand\orbnum{m}

\begin{frontmatter}

  \title{From template analysis\\ to generating partitions II:\\
    Characterization of the symbolic encodings}

\author{Jérôme Plumecoq \and Marc Lefranc\thanksref{CORR}}

\address{Laboratoire de Physique des Lasers, Atomes, Mol\'ecules, UMR
  CNRS 8523,\\
  Centre d'\'Etudes et de Recherches Lasers et Applications,\\
  Universit\'e de Lille I, F-59655 Villeneuve d'Ascq Cedex, France}
\thanks[CORR] {Corresponding author. E-mail: Marc.Lefranc@univ-lille1.fr}

\begin{abstract}
  We give numerical evidence of the validity of a previously described
  algorithm for constructing symbolic encodings of chaotic attractors
  from a template analysis. We verify that the different solutions
  that can be found are dynamically equivalent, and that our approach
  yields results that are consistent with those obtained from methods
  based on homoclinic tangencies. This is further confirmed by
  verifying directly that the computed partitions are generating to a
  high degree of accuracy, and that they can be used to estimate
  precisely the metric entropy. It is also shown that the correct
  number of symbols needed to describe the dynamics is naturally
  provided, and that a compact parameterization of a partition can
  easily be determined, which makes our algorithm suitable for
  applications such as real-time encoding.
\end{abstract}
\begin{keyword}
  Generating partitions. Symbolic Dynamics. Template analysis. Knot theory.\\
  PACS 98: 05.45.+b
\end{keyword}

\end{frontmatter}

\tableofcontents
\newpage

\section{Introduction}
In the first part of this work~\cite{PlumecoqP99a}, we have presented
an algorithm to construct symbolic encodings of a chaotic attractor,
which follows the approach proposed in~\cite{Lefranc94a}. This
algorithm utilizes template
analysis~\cite{Mindlin90a,Mindlin91a,Gilmore98a,Tufillaro92a,Solari96a}
to extract symbolic dynamical information from the topological
invariants of a set of unstable periodic orbits (UPO) embedded in the
attractor.

Template analysis has its roots in the observation that the
topological organization of unstable periodic orbits of a
three-dimensional hyperbolic flow can be studied in a systematic
way~\cite{Birman83a,Ghrist97a}. More precisely, there exists a
branched two-dimensional manifold, a \emph{template}, on which all the
unstable periodic orbits of such a flow can be laid without modifying
their knot-theoretic invariants. The structure of the template thus
describes concisely the global organization of the flow.

Because periodic orbits cannot intersect themselves when a control
parameter is varied, a set of UPO embedded in a three-dimensional
chaotic attractor can be tracked to a regime where the system is
hyperbolic without modifying their invariants.  In the hyperbolic
case, there is a natural coding for periodic orbits: their symbolic
names are obtained by listing which branches are successively visited
by their projections on the template. It is thus only natural to
extend this coding to experimental orbits by requiring that each UPO
is assigned the symbolic name of one of the template orbits that has
the same topological invariants. This correspondence is one-to-one for
a large number of low-period orbits.

Once the possible symbolic names for each detected UPO have been
identified in this way, this information must be combined with the
knowledge of the position of periodic points in a section plane to
construct a generating partition of the latter. In our algorithm,
partitions are parameterized by a set of periodic points associated
with given symbols, called \emph{reference points}. To represent an
arbitrary trajectory by a symbolic sequence, each intersection with
the section plane is encoded by the symbol attached to the closest
reference point. Starting from an initial partition based on a small
set of low-period orbits (for example a period-$1$ and a period-$2$
orbits), our method progressively refines this partition by inserting
higher-period orbits in it while preserving the simplicity of its
structure.

The main result of~\cite{PlumecoqP99a} is that this method allows one
to obtain high-resolution partitions that have a simple structure, yet
which are such that the topological invariants of each detected
periodic orbit can be directly read from the symbolic name it has been
assigned. That this is at all possible is by itself a strong
indication of the validity of the approach. Indeed, we had
successfully analyzed a set of more than 1500 orbits, whose
intertwining was described by several millions of integers. Preserving
the topological consistency of the partition in such a case is by no
means trivial. As a matter of fact, we have observed that our
algorithm quickly concludes that no consistent solution can be
obtained when fed with some arbitrarily chosen initial partitions, or
when the input value of a single invariant is purposely modified. This
can be because no simple template can be found or because it becomes
at some point impossible to make the encoding continuous (such that
neighboring points are associated with close symbolic sequences).  To
be fully convinced of the validity of this approach, however,
additional verifications have to be carried out. The aim of this paper
is to provide such evidence.

A first test is related to the internal consistency of the algorithm.
As different initial partitions lead to different final partitions, we
must verify that the latter describe the same dynamics. Returning to
the test partition we had obtained in~\cite{PlumecoqP99a}, we show in
Sec.~\ref{sec:choice} that this is indeed the case: the partitions
that are obtained from the simplest initial partitions that can be
built with orbits of periods $1$, $2$, and $4$ are images or
preimages of each other.

If partitions obtained by a topological approach are relevant, they
should easily identify cases where more than two symbols are needed to
describe faithfully the dynamics. We show that this is so in
Sec.~\ref{sec:threebranch}, by considering attractors of the modulated
laser equations and of the Duffing equations which display a
three-symbol dynamics. The latter example is particularly interesting
because a generating partition had previously be obtained for it in
Ref.~\cite{Giovannini91a}, using a method based on homoclinic
tangencies~\cite{Grassberger85a}.

Because the latter method, which is supported by strong numerical
evidence~\cite{Grassberger89a,Cvitanovic88a,DAlessandro90a,Giovannini91a},
is completely orthogonal to ours, a direct comparison of the outputs
of both methods is probably the most stringent test one can consider
for assessing the validity of the topological approach. As will be
seen in Sec.~\ref{sec:homtang_check}, the agreement is excellent: a
line of principal homoclinic tangencies is entirely located in the
triangles enclosing the border of the partition that was obtained
in~\cite{PlumecoqP99a} with a precision of $10^{-4}$. This not only
show the equivalence of the two approaches, but also that our
algorithm provides reliable error bounds.

This gives an indirect proof of the fact that the partitions we obtain
are generating, as this is generally believed to be true for methods
based on homoclinic tangencies. Nevertheless, we carry out in
Sec.~\ref{sec:generating_test} a direct test of this property. It
allows us to show that, with an appropriate balance between the
numbers of forward and backward symbols, the largest distance between
points whose symbolic sequences coincide on a large number of symbols
is of the order of the precision with which the partition has been
determined. The dependence of this distance with respect to the number
of coinciding symbols also indicates that even partitions determined
with a relatively low precision suffice to characterize precisely the
symbolic dynamics of a chaotic attractor.

In Sec.~\ref{sec:entropy}, we compare estimates of the metric entropy
obtained (i) from a Lyapunov exponent calculation and (ii) from
probabilities of symbol sequences computed from a partition. Again, we
find an excellent agreement. For reference, we also characterize the
symbolic dynamics of the attractors we have considered in this work by
giving for each of them the list of irreducible forbidden words.

Finally, we discuss in Sec~\ref{sec:pract_app} how to provide a
compact description of a partition by reducing as much as possible the
number of periodic points parameterizing it.  Last, we conclude by
discussing possible extensions and applications of our
work.

\section{Notations}
\label{sec:notations}
We first briefly recall the notations that have been defined
in~\cite{PlumecoqP99a}. The input data fed to the algorithm is a set
$\mathcal{O}$ of $m$ periodic orbits $O_i$, whose intersections with a
section plane are noted $O_i^j$, $j=1,\ldots,p_i$. From the
topological invariants of these orbits, a set of possible symbolic
names $\nameofi{O_i}{k}$ is first determined. These names are written
inside brackets (e.g., $\nameof{O_2}=\orbit{01}$), to distinguish them
from symbolic names of periodic points, which are written using
overlined strings (e.g., $\pnameof{O_2^1}{}=\perpoint{01}$). Indeed,
different cyclic permutations of a given name represent the same
periodic orbit, but different periodic points.

A partition $\pa$ of a section plane divides it in disjoint regions
$\region_i (\pa)$, $i=1,\ldots,n$. The symbolic name
$\pnameof{O_i}{\pa}$ assigned by a partition $\pa$ to a periodic orbit
$O_i$ is defined as being the one assigned to its first periodic
point: $\pnameof{O_i}{\pa}=\pnameof{O_i^1}{\pa}=\perpoint{s_1\ldots
  s_{p_i}}$, which is such that $O_i^k\in \region_{s_k}(\pa)$.

In~\cite{PlumecoqP99a}, we had obtained two partitions of an attractor
of the modulated laser equations. The first was computed from a set of
periodic orbits providing a uniform cover of a Poincar\'e section of
the attractor, with a resolution of $10^{-3}$ (in units of the
attractor width), and will be hereafter noted $\pa_a^f$. This partition
and the corresponding set of orbits will be used in
Sec.~\ref{sec:choice} to study the relation between partitions
computed from different initial partitions, and in
Sec.~\ref{sec:pract_app} to describe how a compact parameterization of
a partition may be obtained.

Using an improved algorithm, and a set of orbits such that the region
of the partition border was covered with a high resolution, we had
also obtained a second partition whose border line was localized with
a precision of $10^{-4}$. This partition will be noted $\pa_h^f$ in the
following. It will be used to carry out precise tests related to the
equivalence with methods based on homoclinic tangencies
(Sec.~\ref{sec:homtang_check}), the closeness to a generating
partition (Sec.~\ref{sec:generating_test}), and estimation of the
metric entropy (Sec.~\ref{sec:entropy}).

\section{Influence of the choice of the initial partition}
\label{sec:choice}

\begin{figure}[htbp]
  \begin{center}
    \leavevmode
    \includegraphics[width=13cm]{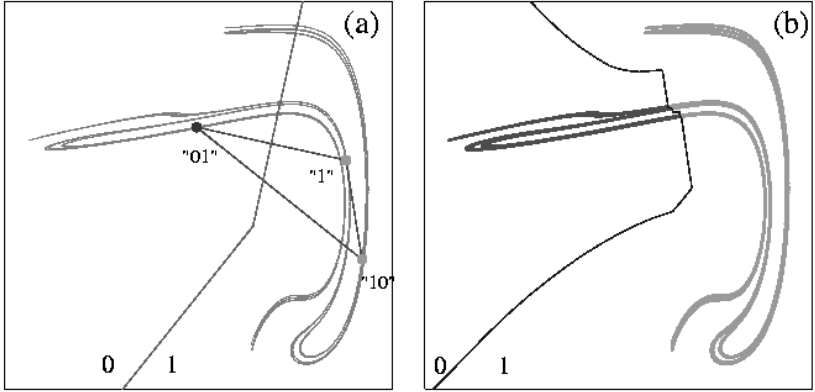}
    \caption{Symbolic encoding of the attractor used as an example in the
      description of the algorithm in~\cite{PlumecoqP99a}. (a) initial
      partition, (b) final partition.}
    \label{fig:ref_part}
  \end{center}
\end{figure}

We have reproduced in Fig.~\ref{fig:ref_part} the partition $\pa_a^f$,
with the initial partition $\pa_a$ that had been used as input.  As
explained in~\cite{PlumecoqP99a}, there is some arbitrariness in the
choice of this initial partition. For instance, we know that one point
of the period-2 orbit $\orbit{01}$ must be assigned the name
$\perpoint{01}$ and the other $\perpoint{10}$, but this gives two
possible choices, which are in principle equivalent:
$\{\pnameof{O_2^1}{\pa}, \pnameof{O_2^1}{\pa}\}$ = $\{\perpoint{01},
\perpoint{10}\} $ (as in Fig.~\ref{fig:ref_part}), or
$\{\pnameof{O_2^1}{\pa}, \pnameof{O_2^1}{\pa}\}$ = $\{\perpoint{10},
\perpoint{01}\} $.  Different initial partitions will obviously lead
to different final partitions, since the algorithm does not modify the
symbols of the reference points after they are inserted in the
partition.

We must therefore verify that the solutions obtained using different
initial partitions model equivalently the symbolic dynamics of the
chaotic attractor, i.e., that the quantities which can be computed
using the partition do not depend on the choice of the initial
partition, provided the latter is reasonably chosen.

Orbits with a unique topological name are not a problem in this
respect.  If they were the only orbits to consider, they could in
principle be named using any cyclic permutation of their topological
name. The first step of the algorithm described in~\cite{PlumecoqP99a}
merely ensures that the permutation we choose for each orbit preserves
the simplicity of the partition. However, the names of the remaining
orbits are determined using not only the topological invariants, but
also an intermediate partition.  These names might thus vary depending
on which initial configuration is chosen, and we must verify that this
is not the case. 

An initial partition is based on $l$ periodic orbits
$O_{i_1},O_{i_2},\ldots,O_{i_l}$ of periods $p_{i_j}$, each having a
unique topological name $\nameof{O_{i_j}}$. Such a partition is
completely defined if one specifies for each orbit $O_{i_j}$ the
symbolic name $\pnameof{O_{i_j}}{\pa} = \pnameof{O_{i_j}^1}{\pa} =
\sigma^{l_j}\nameof{O_{i_j}}$ ($0\leq l_j<p_{i_j}-1$) assigned to its
first periodic point. The initial triangulation then consists of the
reference points $O_{i_j}^k$, associated with the symbols
corresponding to the chosen cyclic permutations.

In the following, we note
$\pa_{i_1,\ldots,i_n}^{\pname_1,\ldots,\pname_l} $ an initial
partition specified in this way, where $N_j=\pnameof{O_{i_j}^1}{\pa}$.
With this notation, $\pa_a=\pa_{1,2}^{\perpoint{1},\perpoint{01}}$
represents the initial partition shown in Fig.~\ref{fig:ref_part}(a)
(with $O_2^1$ being the leftmost periodic point of the $\orbit{01}$
orbit).  

Since our algorithm is deterministic, the final partition
$\pa^f=\pa_{1,\ldots,\orbnum}^{N_1,\ldots,N_\orbnum}$ that contains
all the $\orbnum$ periodic orbits detected in the attractor is a
functional of the initial partition, which we will note:
\begin{equation}
  \label{eq:partfunc}
  \pa^f=A[\pa_{i_1,\ldots,i_l}^{\pname_1,\ldots,\pname_l}] 
\end{equation}

\begin{figure}[htbp]
  \begin{center}
    \leavevmode
    \includegraphics[width=13cm]{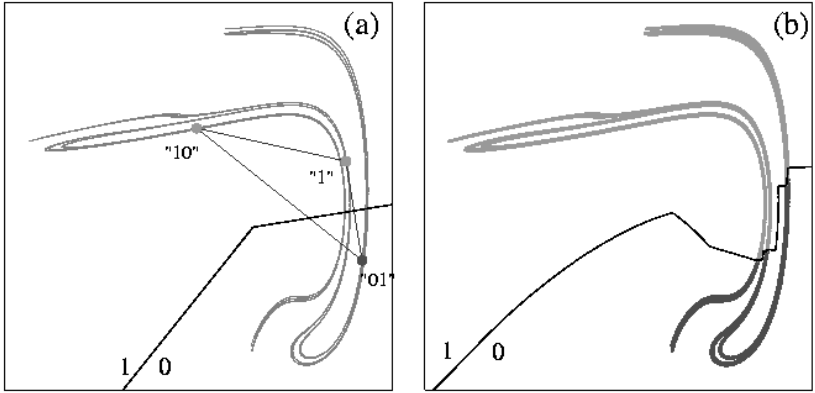}
    \caption{(a) Another possible initial partition using the first two
    orbits. (b) the resulting partition.}
    \label{fig:partinit2}
  \end{center}
\end{figure}

Consider now the second possible initial partition based on the
period-$1$ and period-$2$ orbits, which is
$\pa_b=\pa_{1,2}^{\perpoint{1},\perpoint{10}}$ (shown in
Fig.~\ref{fig:partinit2}a). The difference with $\pa_a$ in
Fig.~\ref{fig:ref_part}a is that the symbols associated with the
points of the period-$2$ orbit have been exchanged.  Using this new
initial partition, we obtain a final partition $\pa^f_b=A[\pa_b]$
(Fig.~\ref{fig:partinit2}b), that markedly differs from
$\pa_a^f=A[\pa_a]$ (Fig.~\ref{fig:ref_part}b).

In fact, $\pa_a^f$ and $\pa_b^f$ are related in a very simple way, and
describe the same dynamics.  Indeed, if we compare for each orbit
$O_i$ the symbolic names $\pnameof{O_i}{\pa_a^f}$ and
$\pnameof{O_i}{\pa_b^f}$ assigned by these two partitions, we find
that in each case the latter is recovered by shifting the former by
one symbol:
\begin{equation}
  \label{eq:shift}
  \forall\, O_i\in \mathcal{O},\quad
  \pnameof{O_i}{\pa^f_b}=
  \sigma\pnameof{O_i}{\pa^f_a}
\end{equation}
For example, $\pnameof{O_3}{\pa_a^f}=\perpoint{0111}$ and
$\pnameof{O_3}{\pa_b^f}=\perpoint{1110}$. As a first result, we note
that the orbits that had several compatible topological names are
identified in the same way by the two final partitions (i.e., up to a
cyclic permutation). This seems to indicate that the symbolic names
which are singled out in the final partition do not result from an
arbitrary choice, but are dynamically relevant.

If we define the action of the shift operator $\sigma$ on a partition
as follows (the symbolic names of the orbits parameterizing the
partition are all shifted by one symbol):
\begin{equation}
  \label{eq:shiftpart}
  \sigma\pa_{i_1,\ldots,i_n}^{N_1,\ldots,N_n}=
  \pa_{i_1,\ldots,i_n}^{\sigma N_1,\ldots,\sigma N_n},
\end{equation}
and since relation (\ref{eq:shift}) is verified for every periodic
orbit involved in the final partitions, we can write:
\begin{equation}
  \label{eq:partshift}
  \pa^f_b =\sigma\pa^f_a
\end{equation}

It is trivial to remark that a similar relation holds for the initial
partitions:
\begin{equation}
  \label{eq:partinitshift}
  \pa_b=\pa_{1,2}^{\perpoint{1},\perpoint{10}}=
  \sigma\pa_{1,2}^{\perpoint{1},\perpoint{01}}=\sigma\pa_a
\end{equation}
and thus we have a kind of commutative relation:
\begin{equation}
  \label{eq:commuting}
  \pa^f_b=A[\sigma\pa_a]=\sigma A[\pa_a]=\sigma\pa^f_a
\end{equation}

While relation (\ref{eq:commuting}) is formally correct in our
particular example, it should not be extrapolated to higher powers of
the shift operator, because $\sigma^2\pa_a=\pa_a$, but
$\sigma^2\pa_a^f\neq\pa_a^f$.  Indeed, for any partition defined with
orbits of periods $p_i$, we have $\sigma^l\pa=\pa$ where $l$ is the
least common multiple of the $p_i$.  What Eq.~(\ref{eq:commuting})
essentially indicates is that our algorithm preserves the dynamical
relation that exists between initial partitions $\pa_a$ and $\pa_b$.
This seems to indicate that it converges to the simplest partition
that is consistent with the information contained in the initial
partition.

Eq.~(\ref{eq:partshift}) also shows that the two solutions obtained
from two different initial conditions are dynamically equivalent, as
is best seen by translating Eq.~(\ref{eq:commuting}) into a relation
involving the first return map $f$.  Since the shift operator is
conjugate to $f$, we have for any detected periodic point $O_i^k \in
\mathcal{P}$:
\begin{equation}
  \label{eq:perpoint}
  \pnameof{O_i^k}{\pa^f_b}=
  \sigma\pnameof{O_i^k}{\pa^f_a}=
  \pnameof{f(O_i^k)}{\pa^f_a},
\end{equation}

Consider the set $\region_i^{\mathcal{P}}({\pa})= \region_i({\pa})
\cap \mathcal{P}$ of all the detected periodic points lying in region
$\region_i({\pa})$ of the partition $\pa$. Eq.~\ref{eq:perpoint}
implies that:
\begin{equation}
  \label{eq:regions}
\forall i,\quad  x \in \region_i^{\mathcal{P}}({\pa_b^f}) \Longleftrightarrow
  f(x)\in \region_i^{\mathcal{P}}({\pa_a^f})
\end{equation}
and therefore:
\begin{equation}
  \label{eq:imageper}
  \forall i,\quad
  \region_i^{\mathcal{P}}({\pa_b^f})=
  f^{-1}\left(\region_i^{\mathcal{P}}({\pa_a^f})\right)
\end{equation}
Because $\region_i^{\mathcal{P}}({\pa_a^f})$ approximates
$\region_i({\pa_a^f})$ (the periodic points were chosen so as to
provide a uniform cover of the attractor), we have:
\begin{equation}
  \label{eq:image}
  \forall i,\quad
  \region_i({\pa_b^f})\approx
  f^{-1}\left(\region_i({\pa_a^f})\right)
\end{equation}
where $\region_i(\pa)$ should be understood as the intersection
of region $i$ with the support of the strange attractor.

Since the image of a partition
$\pa=\{\region_0(\pa),\region_1(\pa),\ldots\}$ can be defined
naturally as $f(\pa) =
\{f(\region_0(\pa)),f(\region_1(\pa)),\ldots\}$, the relation between
partitions $\pa_a^f$ and $\pa_b^f$ can be concisely written as:
\begin{equation}
  \label{eq:borders}
  \pa_b^f \approx f^{-1}\left(\pa_a^f\right)
\end{equation}

The reason why the border of $\pa_b^f$ shown in
Fig.~\ref{fig:partinit2}b is not exactly the preimage of the border of
$\pa_a^f$ shown in Fig.~\ref{fig:ref_part}b is that the mediators of
the triangles in the latter partition need not be exactly mapped to
those of the former. Therefore, if we consider a long chaotic
trajectory, and encode it using the two partitions, the obtained
symbolic sequences will not be entirely shift-equivalent (whereas this
is rigorously true for the periodic orbits used in the partition):
there might be discrepancies for points located very close to the
border.  The degree of equivalence will therefore depend on the
precision with which the partition has been determined, i.e., on the
size of the triangles enclosing the border of the partition.

To assess the degree of equivalence in our example, we have analyzed a
chaotic trajectory of $10^7$ iterations, and compared the symbolic
sequences obtained with the two partitions (taking into account the
shift by one symbol): we found that less than 0.02\% of the symbols
differed.  These observations clearly show that the final partitions
obtained using the two initial partitions describe essentially the
same dynamics. Note also that to localize the border with a higher
precision, it should be possible to use a more sophisticated
interpolation scheme where these discrepancies would be taken into
account to obtain a solution that is consistent with its images and
preimages.  

We have examined all initial partitions based on the period-$1$ and
period-$2$ orbits.  To test further the robustness of our method, we
now consider initial partitions based on the four periodic points of
the period-$4$ orbit, which will allow us to explore different
configurations.

We have four possible initial partitions, each corresponding to a
different cyclic permutation of the topological name
$\nameof{O_3}=\orbit{0111}$. In the first two of these partitions, namely
$\pa_3^{\perpoint{0111}}$ and $\pa_3^{\perpoint{1110}}$, the
period-$4$ orbit is assigned the same name as in the final partitions
$\pa_a^f$ and $\pa_b^f$, respectively. Not surprisingly, we obtain
from these two initial configurations the two final partitions that
have previously been computed: $A[\pa_3^{\perpoint{0111}}] = \pa_a^f =
A[\pa_{1,2}^{\perpoint{1},\perpoint{01}}]$ and
$A[\pa_3^{\perpoint{1110}}] = \pa_b^f =
A[\pa_{1,2}^{\perpoint{1},\perpoint{10}}]$.

\begin{figure}[htbp]
  \begin{center}
    \leavevmode
    \includegraphics[width=12cm]{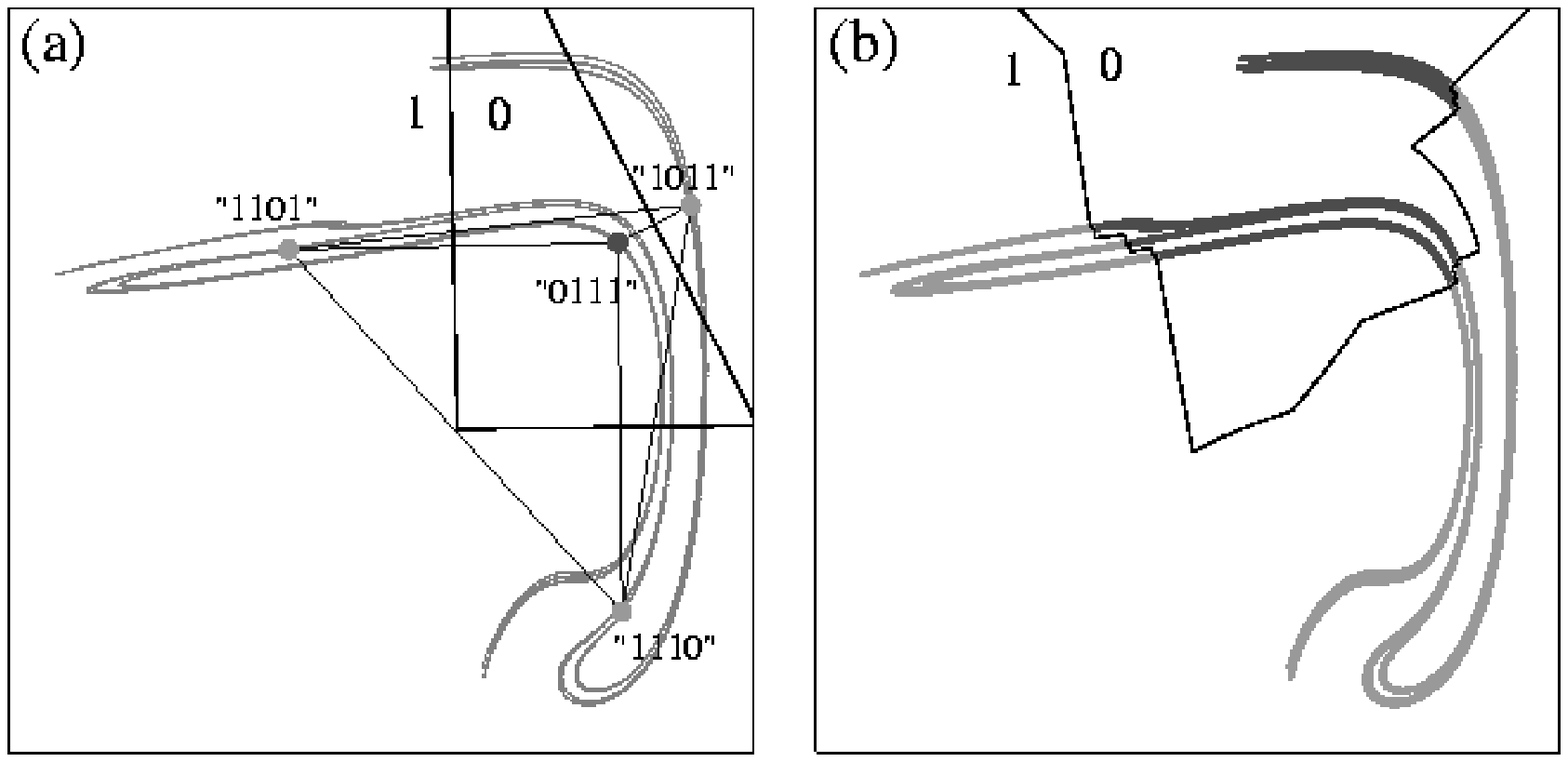}
    \caption{(a) Initial partition $\pa_3^{\perpoint{1101}}$ ; (b) the
      corresponding final partition $A[\pa_3^{\perpoint{1101}}]$} 
    \label{fig:part3}
  \end{center}
\end{figure}

\begin{figure}[htbp]
  \begin{center}
    \leavevmode
    \includegraphics[width=12cm]{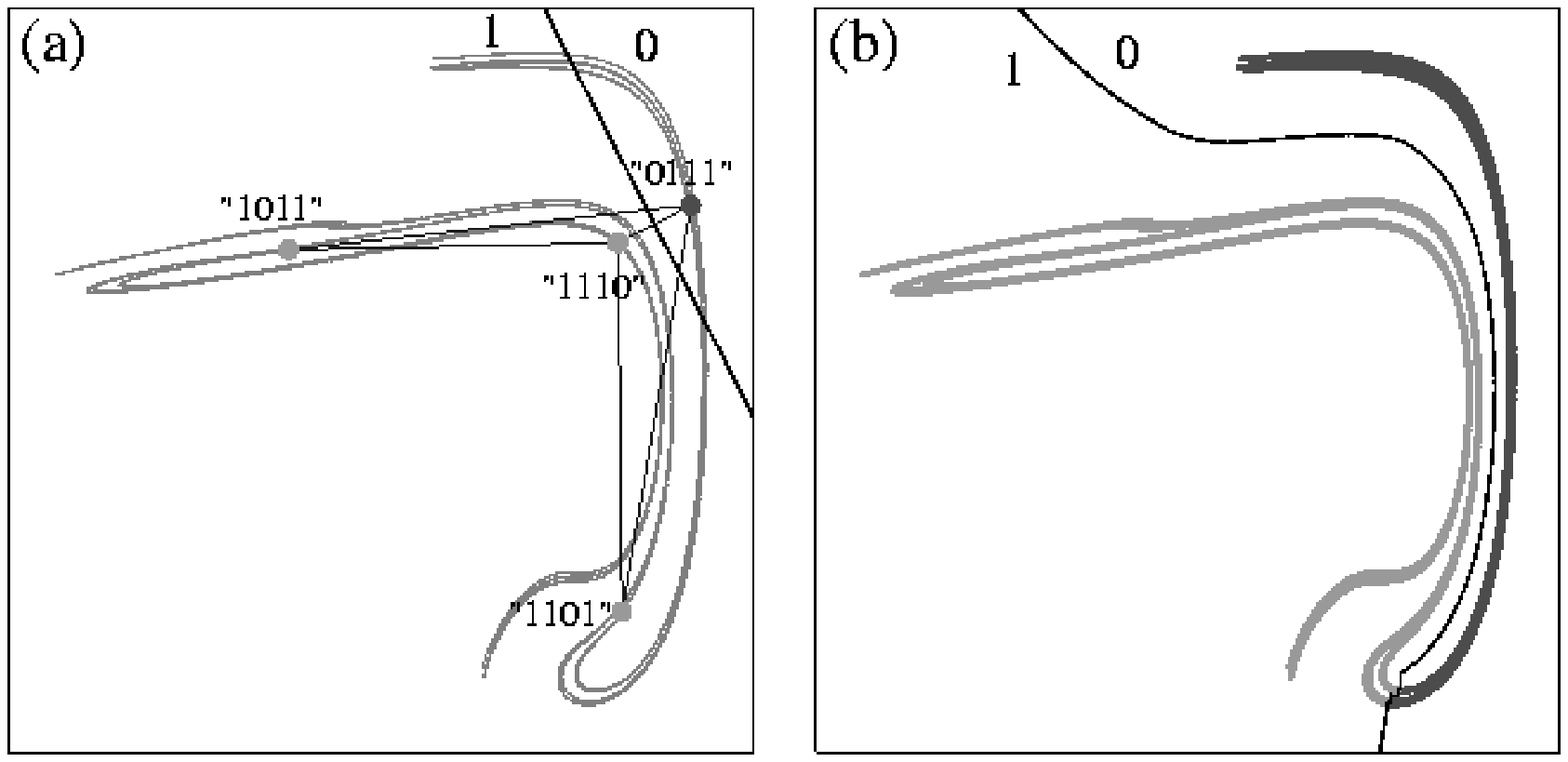}
    \caption{(a) Initial partition $\pa_3^{\perpoint{1011}}$ ; (b) the
      corresponding final partition $A[\pa_3^{\perpoint{1011}}]$} 
    \label{fig:part4}
  \end{center}
\end{figure}

The two other initial partitions, $\pa_3^{\perpoint{1101}}$ and
$\pa_3^{\perpoint{1011}}$, shown in Figs~\ref{fig:part3}a and
\ref{fig:part4}a, yield different partitions which are displayed in
Figs~\ref{fig:part3}b and \ref{fig:part4}b, respectively.

Again, these partitions are equivalent to the partition $\pa^f_a$
shown in Fig.~\ref{fig:ref_part}b, since we have
$A[\pa_3^{\perpoint{1101}}] = \sigma^{2} \pa_a^f$, and
$A[\pa_3^{\perpoint{1011}}] = \sigma^{-1} \pa_a^f$.

The four partitions based on the period-$4$ orbit are therefore
related by:
\begin{equation}
  \label{eq_4t_shift}
  \sigma^{-2} A[\pa_3^{\perpoint{1101}}] = \sigma^{-1}
 A[\pa_3^{\perpoint{1110}}] = 
 A[\pa_3^{\perpoint{0111}}] = \sigma A[\pa_3^{\perpoint{1011}}]
\end{equation}
which translates to:
\begin{equation}
  \label{eq_4t_equiv}
  f^{-1}\left(A[\pa_3^{\perpoint{1011}}]\right) \approx
  A[\pa_3^{\perpoint{0111}}]\approx
  f\left(A[\pa_3^{\perpoint{1110}}]\right) \approx
  f^2\left(A[\pa_3^{\perpoint{1101}}]\right)
\end{equation}

which clearly shows that the six simplest initial configurations (two
based on $\{O_1,O_2\}$ and four on $O_3$) lead to 4 dynamically
equivalent final partitions.

However, it should be noted that our algorithm does not converge for
every possible initial partition. For example, we have tried the 8
initial partitions based on the 7 periodic points of the set of orbits
$\{O_1,O_2,O_3\}$. Four or them lead to the four final partitions that
have already been obtained: these are the initial partitions of the
form
$\sigma^k\pa_{1,2,3}^{\perpoint{1},\perpoint{01},\perpoint{0111}}$.
The four other initial partitions are of the form
$\sigma^k\pa_{1,2,3}^{\perpoint{1},\perpoint{01},\perpoint{1110}}$: in
this case, the algorithm terminates after finding an inconsistency, or
yields a partition which has a complicated structure. Note that the
latter four initial partitions do not belong to the $\sigma$-orbit of
the $\pa_{1,2,3}^{\perpoint{1},\perpoint{01},\perpoint{0111}}$ initial
partition which leads to the simple solution $\pa_a^f$ shown in
Fig.~\ref{fig:ref_part}b.

This observation and other tests we carried out lead us to conclude
that our algorithm is successful whenever the input names defining the
initial partition are those assigned by a partition that is one of the
first few (pre-) images of the $\pa_a^f$ partition.  This supports the
idea that there is a natural solution of lowest complexity, and that
our algorithm successfully approximates it.

To conclude this section, we note an interesting consequence of the
fact that our partitions are parameterized by periodic orbits: forward
and backward images of partitions can easily be computed without being
faced with the exponential divergence of nearby orbits. To do this, it
suffices to shift all the symbolic names given by the final partition
by the same amount, and to recompute the border. This property could
be useful in chaos-based digital communications: by using the
$\sigma^n\pa_a^f$ partition, one can predict $n$ iterations in advance
the symbol corresponding to the standard partition $\pa_a^f$, allowing
for early error recovery.

\section{Examples of three-symbol partitions}
\label{sec:threebranch}

A difficult problem when trying to construct a symbolic encoding of a
given attractor is to determine how many different symbols are needed
to describe faithfully the dynamics. If topological entropy cannot be
easily computed, metric entropy, which is a lower bound of topological
entropy, could be estimated from the greatest Lyapunov exponent.  For
example, at least $n$ symbols should be used when $\lambda_1T>\ln n$,
where $T$ is the modulation period. However, there can be situations
where the right number of symbols is larger than indicated by the
topological entropy because there are many forbidden symbolic
sequences.

Besides their robustness, methods based on template analysis have the
advantage that the correct number of symbols is automatically obtained
from the preliminary template analysis: this is just the number
of branches of the simplest template that describes the topological
organization of the data.

To illustrate this, we now apply our algorithm to two strange
attractors whose symbolic dynamics involve three symbols: an attractor
of the modulated laser equations used in~\cite{PlumecoqP99a}, but at
different parameter values, and an attractor of the Duffing system at
the parameters studied in Ref.~\cite{Giovannini91a}, where a
generating partition was determined from the locations of the
homoclinic tangencies.

\subsection{Symbolic encoding of a  spiral modulated laser
  attractor}
\label{sec:laser_spiral}
Depending on the parameters, the regimes displayed by a chaotic system
can have different degrees of complexity: some may correspond to a
two-symbol dynamics while the description of other, more chaotic,
regimes requires a larger number of symbols.

This is the case of the modulated laser equations whose parameter
space is divided in several regions corresponding to different
topological organizations, as is common for nonlinear
oscillators~\cite{Gilmore95a}. To test our algorithm in a more
complicated case than the one considered in~\cite{PlumecoqP99a}, we
have studied an attractor close to the three-branch spiral attractor
described in Ref.~\cite{Boulant98a}, and whose parameters are the same
as in~\cite{PlumecoqP99a}, except that $m=0.08125$, $\gamma=1.5\times
10^{-3}$, and $\omega\approx 0.923077 \omega_r$ where
$\omega_r=\sqrt{\gamma(A-1)}$ is the relaxation frequency of the
laser.

\begin{figure}[htbp]
  \begin{center}
    \leavevmode \includegraphics[width=10cm]{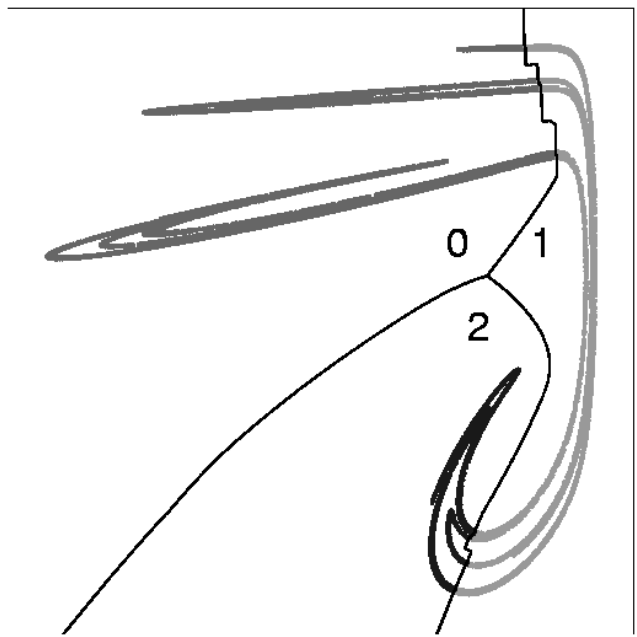}
    \caption{Generating partition for a three-branch spiral
      attractor of the modulated laser equations.}
    \label{fig:spiral_part}
  \end{center}
\end{figure}

That an analysis of this attractor requires at least three symbols is
readily indicated by the fact that it contains four period-$3$ orbits,
whereas at most two exist in a two-symbol dynamics. This is confirmed
by a template analysis, which shows that the topological structure of
this attractor is described by a spiral three-branch template (a.k.a.
``G\^ateau Roul\'e''), described by the following template and
layering matrices:
\begin{equation}
  \label{spiral_template}
    t^{S} =
  \left(
    \begin{array}[c]{ccc}
      0 & 0 & 0\\
      0 & 1 & 2\\
      0 & 2 & 2
    \end{array}
  \right),\quad
  {l}^{S} =
  \left(
    \begin{array}[c]{ccc}
      0 & 2 & 1
    \end{array}
  \right)
\end{equation}
An illustration of the corresponding three-branch manifold can be
found in Fig. 6 of Ref.~\cite{Boulant98a}.  The partition shown in
Fig.~\ref{fig:spiral_part} has been obtained using the improved
procedure outlined at the end of Ref.~\cite{PlumecoqP99a}, and is
based on a set of $1250$ orbits of periods up to $32$ which provides a
resolution better than $10^{-3}$ in the region of the border. It
should be noted that in this case, there is only one initial partition
based on the period-$1$ and period-$2$ orbits $\orbit{1}$ and
$\orbit{01}$ that leads to a simple solution, the one which assigns
the symbol ``0'' to the leftmost period-$2$ point.  Indeed, the
algorithm does not converge if we exchange the symbols of the
period-$2$ points.  This is due to the template having three branches:
exchanging these two periodic points would amount to placing branch
``0'' between branches ``1'' and ``2'' and would violate the
requirement of continuity.

\subsection{Symbolic encoding of a Duffing attractor}
\label{sec:duffing}
The Duffing system
\begin{equation}
  \label{eq:duffing_eq}
  \dot{x} = y, \quad \dot{y} = -\delta y +x -x^3+\gamma \cos f t
\end{equation}
is one of the few three-dimensional flows for which a generating
partition has been determined by methods based on homoclinic
tangencies~\cite{Giovannini91a}. To verify that our approach yields
results that agree with these methods, we have therefore applied our
procedure to the regime studied in Ref.~\cite{Giovannini91a}, which
corresponds to parameters $\delta=0.25$, $\gamma=0.4$, and $f=1$.

A direct template analysis indicates that the topological structure
can be described by the following six-branch matrices:

\begin{equation}
  \label{duffing_6b_template}
    t^{D} =
  \left(
    \begin{array}[c]{cccccc}
      2 & 2 & 0 & 0 & 0 & 2 \\
      2 & 1 & 0 & 0 & 0 & 2\\
      0 & 0 & 0 & 0 & 0 & 2\\
      0 & 0 & 0 & 2 & 2 & 2\\
      0 & 0 & 0 & 2 & 1 & 2\\
      2 & 2 & 2 & 2 & 2 & 2
    \end{array}
  \right),\quad
  {l}^{D} =
  \left(
    \begin{array}[c]{rrrrrr}
      0 & -1 & 1 & 3 & 2 & -2
    \end{array}
  \right)
\end{equation}
It can be seen that the template matrix in
Eq.~(\ref{duffing_6b_template}) does not describe a continuous
branched manifold because the torsions of two adjacent branches differ
by two units. Presumably, these matrices correspond to a part of a
larger template. This is however not important, as it turns out that
the dynamics of this regime can be studied with only three symbols.

Indeed, as noted in Refs.~\cite{Giovannini91a,Gilmore95a}, we can take
advantage of a symmetry of the Duffing equations (\ref{eq:duffing_eq})
to simplify the analysis, namely the invariance under the
transformation $x \rightarrow -x, y \rightarrow -y, \phi \rightarrow
\phi + \pi$. This symmetry indicates that the dynamics during the
first half-period and during the second half-period are seentially
equivalent. As a consequence, the Poincaré sections at $\phi=0$ and
$\phi=\pi$ are identical modulo an inversion around the origin. We
will thus consider the reduced dynamical system, where the integration
of Eqs.~(\ref{eq:duffing_eq}) over half a period is followed by a
twist of a half-turn.

If we discard two solutions where torsions of adjacent branches differ
by more than one half-turn and which are algebraically equivalent to
the solutions presented below~\cite{Boulant98a}, we find two possible
templates for this reduced system. The first one is described by the
matrices:

\begin{equation}
  \label{eq:duffing_template}
    t^{D_1} =
  \left(
    \begin{array}[c]{ccc}
      1 & 2 & 2\\
      2 & 2 & 2\\
      2 & 2 & 3
    \end{array}
  \right),\quad
  {l}^{D_1} =
  \left(
    \begin{array}[c]{rrr}
      0 & -2 & -1
    \end{array}
  \right)
\end{equation}
which correspond to the spiral template described by
Eqs.~(\ref{spiral_template}) with an additional half-twist and which
was identified as a building block for the Duffing template by Gilmore
and McCallum~\cite{Gilmore95a}. The other one is a S-shaped template,
whose matrices are:
\begin{equation}
  \label{eq:duffing_template_alt}
    t^{D_2} =
  \left(
    \begin{array}[c]{ccc}
      2 & 2 & 2\\
      2 & 1 & 2\\
      2 & 2 & 2
    \end{array}
  \right),\quad
  {l}^{D_2} =
  \left(
    \begin{array}[c]{rrr}
      0 & -1 & -2
    \end{array}
  \right)
\end{equation}

These two solutions are algebraically compatible with the topological
invariants of the UPO: for each of the two templates, it is possible
to find a set of periodic orbits lying on it that has exactly the same
invariants as the experimental ones. This means that the spectra of
periodic orbits of the two templates have subsets that are isotopic,
and that the set of detected orbits in the Duffing attractor belongs
to this intersection.  Yet, this indeterminacy is intriguing, as the
above two solutions have quite different topological structures.

However, there is a clear-cut difference between these two solutions
which is unveiled when trying to construct a symbolic encoding.
Whereas a partition is easily found using the first template
(described by Eqs.~(\ref{eq:duffing_template})), our algorithm quickly
stops when the second one is used as input: at some point the current
partition is such that some orbits do not any longer have at least one
topological name compatible with it. This means that one cannot find a
symbolic encoding that is simultaneously continuous in the section
plane and reproduces the symbolic dynamics associated with the
template (\ref{eq:duffing_template_alt}). We believe that this
indicates that only the first solution is dynamically relevant. Thus,
although we focus here on the application of template analysis to the
construction of symbolic encodings, we see that our algorithm can also
be helpful to discriminate algebraic solutions of the template-finding
problem.

\begin{figure}[htbp]
  \begin{center}
    \leavevmode
    \includegraphics[width=10cm]{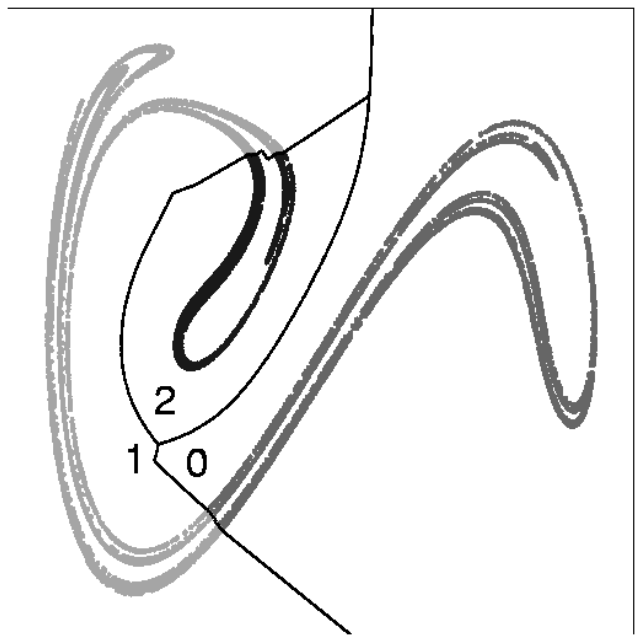}
    \caption{Generating partition obtained for the Duffing attractor.}
    \label{fig:duffing_part}
  \end{center}
\end{figure}

The partition obtained with template (\ref{eq:duffing_template}) is
shown in Fig.~\ref{fig:duffing_part}. It is based on a set of $1326$
periodic orbits of periods up to $36$, providing a cover with a
resolution better than $10^{-3}$ in the neighborhood of the border. It
is seen to reproduce Fig.~4 of Ref.~\cite{Giovannini91a}, where the
partition had been computed by following lines of homoclinic
tangencies. It should be stressed that in our case, this partition is
the simplest and most natural solution (obtained from an initial
configuration based on the period-1 and period-2 orbits). In contrast
with this, it was noted in Ref.~\cite{Giovannini91a} that some
homoclinic tangencies involved in this partition are not primary, and
that heuristic considerations had to be used to select these
particular tangencies.  This shows that the topological approach
provides us with additional information that cannot be obtained from
the study of homoclinic tangencies, and is thus a powerful tool for
selecting the correct lines of tangencies.

Although the partition shown in Fig.~\ref{fig:duffing_part} does not
differ to the eye from the partition computed in
Ref.~\cite{Giovannini91a}, we still have to verify precisely that the
two approaches still agree at the smallest scales.  In particular, we
must check whether the error bounds naturally provided by our
algorithm are correct: if so, the partition border determined using
homoclinic tangencies should be entirely located inside the region
where the coding is considered uncertain (i.e., in the circumcircles
of the triangles with different symbols, see~\cite{PlumecoqP99a}).  We
carry out this verification in the next section.

\section{Comparison with methods based on homoclinic tangencies}
\label{sec:homtang_check}

A difficult problem associated with using homoclinic tangencies to
construct partitions is that there is a countable infinity of
homoclinic tangencies: all the forward and backward images of a
homoclinic tangency are themselves homoclinic tangencies. As a
consequence, there are homoclinic tangencies everywhere in the
attractor (just as the critical point of the logistic map has
preimages in every part of the interval).

However, there are homoclinic tangencies that are more relevant for a
large-scale characterization of an attractor, because their influence
is felt over a wider region: they should be preferably selected to
build a partition line. A possible selection criterion proposed in
Ref.~\cite{Giovannini91a} is that the sum of the curvatures of the
stable and unstable manifolds at the tangency should be minimal,
because the curvature of the stable (unstable) manifold diverges when
following the backward (forward) orbit of a given tangency.

\begin{figure}[htbp]
  \begin{center}
    \leavevmode
    \includegraphics[width=10cm]{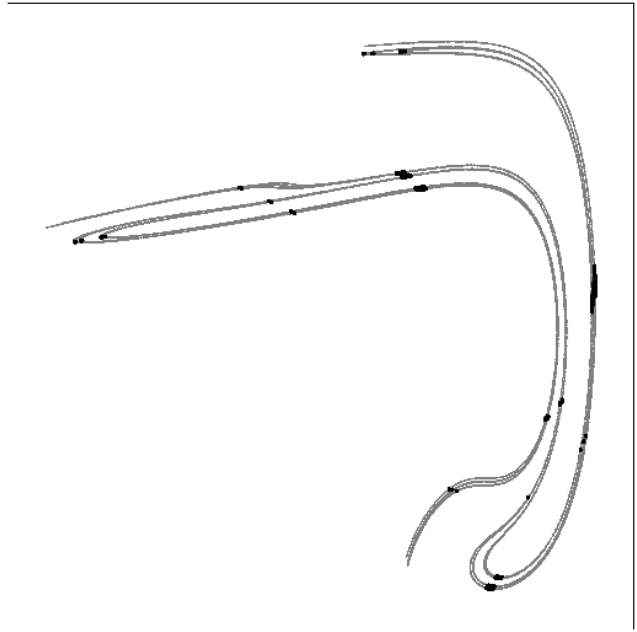}
    \caption{Poincaré section of an attractor of the modulated laser
      equations. The points where the angle between the stable and the
      unstable manifolds is smaller than 0.1 radians are plotted in
      black. }
    \label{fig:main_tang}
  \end{center}
\end{figure}

To illustrate the existence of ``principal'' homoclinic tangencies in
the case of the modulated laser equations (as the same parameters as
in~\cite{PlumecoqP99a}), we have plotted in Fig.~\ref{fig:main_tang}
the points (extracted from a sequence of $10^{8}$ iterations) where
the angle between the invariant manifolds, computed using the method
described in Appendix B of Ref.~\cite{Jaeger97a}, is smaller than a
prescribed value.  It is readily seen that there are regions where
this angle remains small over a significantly large region of the
section plane, whereas no point with a small angle is detected
elsewhere (although there are regions with such points everywhere,
there are too small to be visited in a reasonable amount of time by a
typical trajectory).

A first interesting result is that the wider small-angle regions in
Fig.~\ref{fig:main_tang} are seen to correspond to some of the
partitions that have been obtained in Sec.~\ref{sec:choice} for
different initial configurations. This is obviously related to the
fact that these borders have been shown to be images or preimages of
each other.

\begin{figure}[htbp]
  \begin{center}
    \leavevmode
    \includegraphics[angle=270,width=13cm]{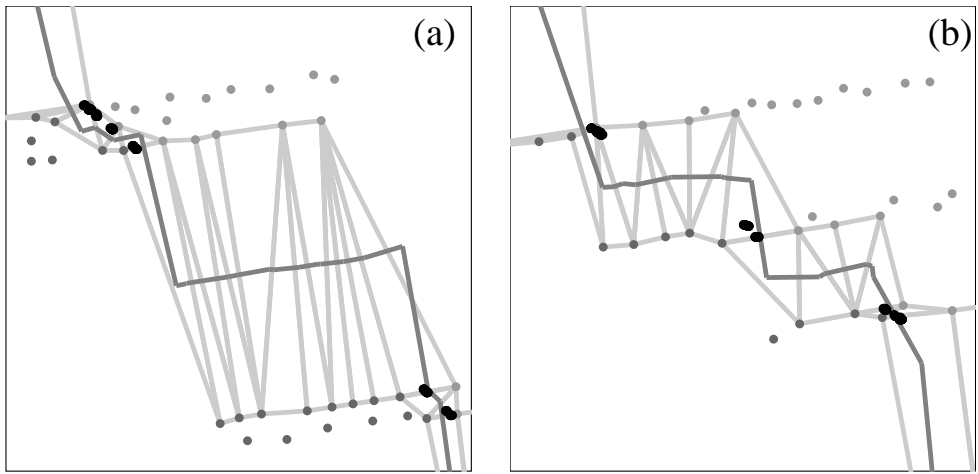}
    \caption{Enlarged views of two neighborhoods of the border line of 
      the high-resolution partition $\pa_h^f$. The size of the two
      represented squares is $2.5\times10^{-3}$, in units of the
      attractor width.  Light dots represent periodic points
      parameterizing the partition.  Heavy dots indicate homoclinic
      tangencies (the angle between the two invariant manifolds at
      these points is below $2\times10^{-4}$ radians).  The border
      line and the triangles enclosing it are also represented. The
      linewidth used to draw the border line is $2.5\times10^{-5}$. }
    \label{fig:homtan_co2_zoom}
  \end{center}
\end{figure}

To make the comparison more precise, we show in
Fig.~\ref{fig:homtan_co2_zoom} enlarged views of neighborhoods of the
border of the high-resolution partition $\pa_h^f$ (see
Sec.~\ref{sec:notations}), in which we have indicated the approximate
locations of homoclinic tangencies. The agreement is seen to be
excellent: all the homoclinic tangencies are located inside the
triangles that enclose the border of the partition.  This does not
only show that the partition line is well approximated by our method,
but also that the error bounds it provides strictly hold.

\begin{figure}[htbp]
  \begin{center}
    \leavevmode
    \includegraphics[angle=270,width=13cm]{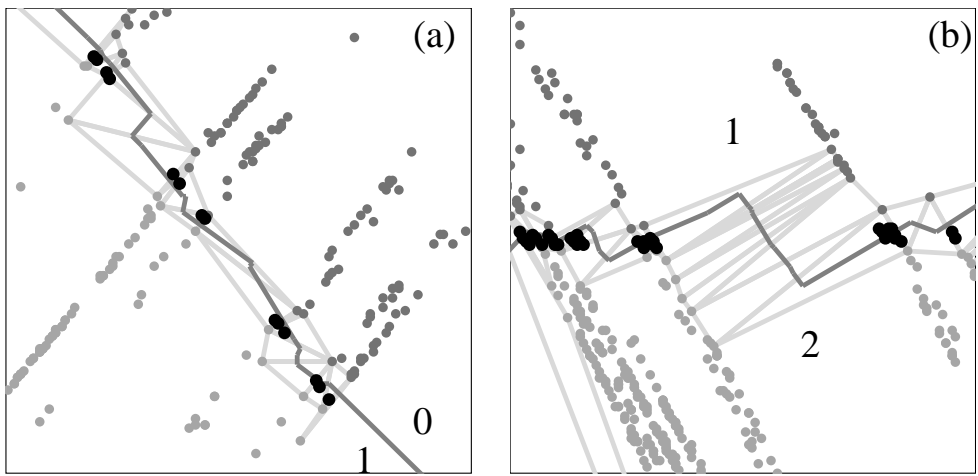}
    \caption{Enlarged views of two neighborhoods of the border line of
      the partition of the Duffing attractor shown in
      Fig.~\ref{fig:duffing_part}. Black heavy dots represent points
      of homoclinic tangency. The widths of the represented squares
      and of the border line are $10^{-2}$ and $10^{-4}$,
      respectively. (a) part of the border between $\region_0$ and
      $\region_1$, (b) part of the border between $\region_1$ and
      $\region_2$. }
    \label{fig:homtan_duffing_zoom}
  \end{center}
\end{figure}

We have also verified this to be true in the case of the spiral laser
attractor and of the Duffing attractor which were considered in
Sec.~\ref{sec:threebranch}. Because the latter example is one of the
few three-dimensional flows for which a generating partition has been
obtained from homoclinic tangencies~\cite{Giovannini91a}, the
equivalent of Fig.~\ref{fig:homtan_co2_zoom} for the Duffing attractor
is displayed in Fig.~\ref{fig:homtan_duffing_zoom}. The reason why the
locations of homoclinic tangencies are better resolved in
Fig.~\ref{fig:homtan_duffing_zoom}a than in
Fig.~\ref{fig:homtan_duffing_zoom}b is that the same limit angle
($5\times 10 ^{-4}$ radians) was used in both cases. As the line of
homoclinic tangencies shown in Fig.~\ref{fig:homtan_duffing_zoom}b is
a principal line, but not the one of
Fig.~\ref{fig:homtan_duffing_zoom}a, the region in
Fig.~\ref{fig:homtan_duffing_zoom}b where the angle between invariant
manifolds is smaller than this value is markedly wider than in
Fig.~\ref{fig:homtan_duffing_zoom}a. However, the relatively lower
precision in Fig.~\ref{fig:homtan_duffing_zoom}b still allows one to
verify that homoclinic tangencies are located inside the triangles
enclosing the border of the partition.

As can be seen in Figs.~\ref{fig:homtan_co2_zoom} and
\ref{fig:homtan_duffing_zoom}, the results of this section give strong
evidence that algorithms based on template analysis and on the
structure of the homoclinic tangencies converge to the same answer.
The theoretical tools which these two methods rely on are so utterly
different that this agreement is quite fascinating and shows that
chaotic dynamics can be studied from multiple points of view. Because
of the (at least superficially) complete independence of the two
approaches, we believe that this result not only supports the validity
of the topological approach, but also provides an additional
confirmation of the correctness of methods based on homoclinic
tangencies.

Since partitions based on homoclinic tangencies are believed to be
generating, this should also be true for the partitions obtained with
our algorithm. Nevertheless, in order to make our study complete and
to obtain an estimate of the resolutions that would actually be
required for practical applications, we carry out in the next section
a direct test of this property.

\section{Are topological partitions generating?}
\label{sec:generating_test}

The defining property of a generating partition is that the
trajectories of two different points are encoded with different symbol
sequences.  In our algorithm, this is by construction enforced for the
periodic orbits detected in the strange attractor and utilized in the
topological analysis.  Nevertheless, it is important to quantify the
deviation of the obtained partitions from the ideal case where the
mapping from bi-infinite symbolic sequences to chaotic trajectories is
well-defined and continuous.

More precisely, let us consider ``cylinders'':
\begin{equation}
  \label{eq_cylinders}
  C (s_{-m}\ldots s_{-1}.s_0 s_1 \ldots s_{n-1}) =
  \left\{
    x \,; s_\pa (f^i(x)) = s_i, i=-m,\ldots,n-1
    \right\}
\end{equation}
which are regions of the Poincar\'e section of the attractor
containing points whose $m$ backward and $n$ forward symbols are
identical. A partition $\pa$ is generating if and only if:
\begin{equation}
  \lim_{n,m\to\infty} \delta_M (\pa,m,n) = 0,\quad\text{with }
  \delta_M (\pa,m,n) = \max_{C\in\,\mathcal{C}_{\pa}[m,n]} \delta (C)
\end{equation}
where $\mathcal{C}_{\pa}[m,n]$ is the set of all cylinders with $m$
backward and $n$ forward symbols and $\delta(C)$ is the diameter of a
cylinder.

For a given sequence length $l=m+n$, there is a number $n$ of forward
symbols that will minimize $\delta_M (\pa,l-n,n)$.  Because backward
(resp. forward) symbols specify the position of a point along a
segment of the local stable (resp.  unstable) manifold, the optimal
$n$ and $m$ verify $n\lambda_1\approx m\lambda_2$, with $\lambda_1$
and $\lambda_2$ being respectively the positive and negative Lyapunov
exponents of the return map.

We can thus characterize how close a partition $\pa$ is to a
generating partition by the quantities:
\begin{equation}
  \label{eq_cyl_gen}
  \rho (\pa,l) = \min_n \delta_M (\pa,l-n,n),\quad
  \rho (\pa) = \lim_{l\to\infty} \rho (\pa,l)
\end{equation}
For a generating partition, $\rho (\pa)$ should be zero. If a finite
value is obtained, it quantifies the deviation of the partition $\pa$
from a generating one.

\begin{figure}[htbp]
  \begin{center}
    \leavevmode
    \includegraphics[angle=270,width=12cm]{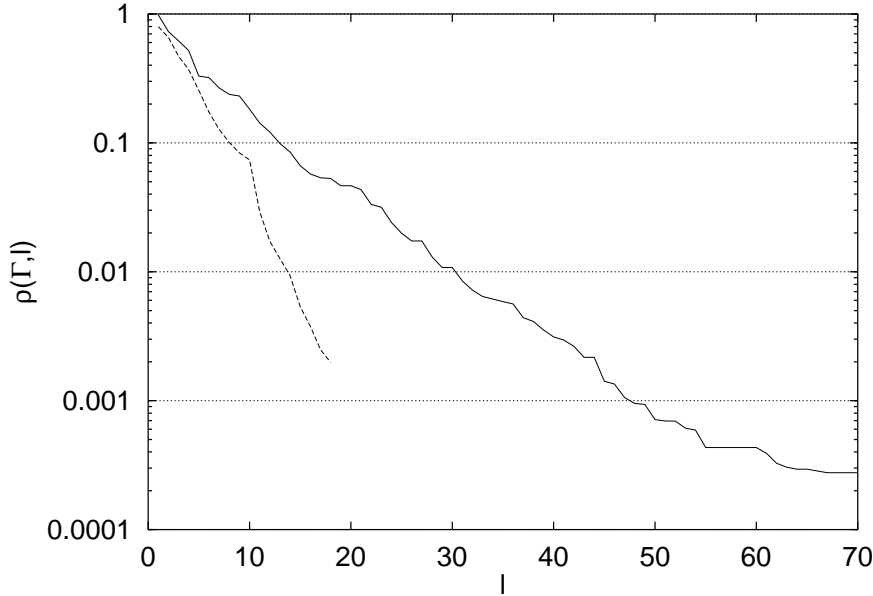}
    \caption{Convergence of $\rho(\pa,l)$ vs. the total
      number of symbols $l$. Diameters are expressed in units of the
      attractor width. For comparison purposes, the dashed line
      represents the evolution of the geometric average of all
      cylinder diameters for a given number of symbols (only values
      for which all cylinder sizes could be reliably estimated are
      shown). The latter curve is seen to converge to zero much more
      rapidly. We have verified that it is almost insensitive to the
      precision with which the border is located.}
    \label{fig:gen_test}
  \end{center}
\end{figure}

We have studied the convergence of $\rho (\pa,l)$ with increasing $l$
for the high-resolution partition $\pa_h^f$ (see
Sec.~\ref{sec:notations}), as shown in Fig.~\ref{fig:gen_test}. After
an approximately logarithmic decrease, $\rho (\pa,l)$ is seen to
saturate slightly below $3\times 10^{-4}$ of the attractor width. It
should be noted however that because $\rho(\pa,l)$ can be reliably
estimated only if there is a sufficient number of points in the
largest cylinder, we did not carry out the computations for $l>70$.
Thus, we cannot guarantee that $\rho(\pa,70)$ represents the final
plateau.  Nevertheless, it is consistent with the fact that the border
of this partition is localized almost everywhere with a precision
better than $10^{-4}$.

\begin{figure}[htbp]
  \begin{center}
    \leavevmode
    \includegraphics[angle=270,width=12cm]{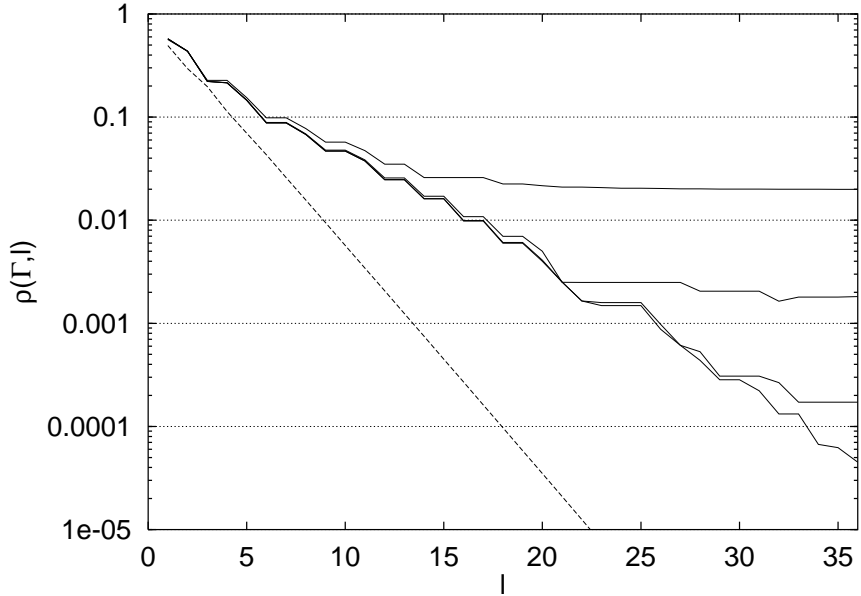}
    \caption{Equivalent of Fig.~\ref{fig:gen_test} for the logistic
      map. The solid curves represent the evolution of the largest
      cylinder diameter vs. the number of symbols for the correct
      partition and for partitions where the border is displaced by a
      fraction $10^{-4}$, $10^{-3}$, and $10^{-2}$ of the attractor
      width.  These curves can be seen to saturate at a value which is
      roughly twice the error on the location of the partition border.
      The dashed line corresponds to the geometric average of cylinder
      diameters. 
      }
    \label{fig:gen_test_logis}
  \end{center}
\end{figure}

To illustrate the fact that the saturation value of the quantity
(\ref{eq_cyl_gen}) provides an estimate of the quality of the
partition, we have performed the same test for the logistic map
$x_{n+1}=4 a x_n (1-x_n)$ at $a=0.955$, where it has roughly the same
symbolic dynamics as our sample system (up to length 12, there is only
one forbidden sequence, which is ``00''). Besides the correct
partition, whose border is located at the critical point, we
characterized different partitions with borders displaced by a small
amount (Fig.~\ref{fig:gen_test_logis}). As a rule of thumb,
$\rho(\pa,l)$ appears to saturate at roughly twice the error on the
location of the partition border. As this ratio is likely to be larger
in the case of a two-dimensional system, the curve in
Fig.~\ref{fig:gen_test} seems to be consistent with an error on the
partition border of the order of $10^{-4}$ of the attractor width.

A comparison of Figs.~\ref{fig:gen_test} and \ref{fig:gen_test_logis}
also shows that to reach a given maximal diameter, the two-dimensional
case requires roughly twice as many symbols as the one-dimensional
one. We attribute this to the fact that two directions are needed to
locate a point in the former case. As a result, the convergence of the
curve in Fig.~\ref{fig:gen_test} is very slow: it takes 31 symbols for
the maximal diameter to fall below $0.01$. Since saturation is reached
at much smaller scales, this should be a genuine property of the
system and not an artifact of our algorithm.

This slow convergence can be traced back to the nonhyperbolicity of
the attractor. Near the border, hence near homoclinic tangencies, the
stretching rate is very close to 1, and very little information is
gained at each iteration of the return map. This has important
consequences, as this indicates that it would be pointless to require
that the border be localized with a high precision.

Whether we want to characterize the grammar of chaos, i.e., determine
the list of forbidden sequences, or use the chaotic system to transmit
a digital signal~\cite{Hayes93a,Hayes94a}, 10--12 symbols should
represent a reasonable amount of information. From the curve in
Fig.~\ref{fig:gen_test}, we see that the maximal diameter for 12
symbols is around $0.12$. If error estimates obtained from saturation
values can be trusted (Fig.~\ref{fig:gen_test_logis}), a precision of
a few per cent should therefore be sufficient.

In regions of parameter space where the topological entropy is
sufficiently large (the number of forbidden symbolic sequences, hence
of pruned periodic orbits, should not be too large), such a precision
can be easily obtained from a topological analysis of unstable
periodic orbits embedded in an experimental chaotic attractor. In
fact, a precision of about one per cent has already been achieved for
a partition of an attractor observed in a modulated
laser~\cite{Lefranc94a}. Topological methods for constructing symbolic
encodings thus appear to be perfectly suited to the symbolic dynamical
characterization of experimental systems, especially given their
robustness to noise.

To conclude this section, we note that the above remarks are
consistent with the observations reported by Bollt, Lai and
Grebogi~\cite{Bollt97a} in their work on noise resistance in
chaos-based communication schemes. To prevent noise-induced
``bit-flipping'', the trajectory of a chaotic communication device
should be kept at a finite distance of the partition border. Enforcing
a forbidden zone inside the attractor discards some symbolic sequences
and in principle reduces the channel capacity (measured by
the topological entropy). The authors of Ref.~\cite{Bollt97a} noted
that in fact the latter decreases very slowly when the gap is
increased. This is of course due to the near-border cylinders having a
large diameter. In contrast with this, the diameters of cylinders in
the remaining of the attractor decrease very rapidly with the number
of symbols, as is shown by the evolution of the geometric average of
the cylinder diameters in Figs.~\ref{fig:gen_test} and
\ref{fig:gen_test_logis}. This ensures that a system can be forced to
follow a prescribed symbolic sequence using only tiny perturbations.

A key property of a generating partition is that it can be utilized to
estimate various invariant quantities such as metric entropy. In the
following section we characterize our partitions by measuring the
latter and by giving for each of our sample attractors the list of
forbidden sequences.

\section{Estimates of metric entropy}
\label{sec:entropy}

There are quantitative measures of chaos that can be recovered both
from a study of symbolic dynamics and from a direct analysis of
trajectories in phase space. The most important is probably the metric
entropy $h_1$ which can be defined in symbolic dynamical terms as:
\begin{equation}
  \label{eq_entropy}
  h_1 = \lim_{l\to\infty} h_1(l),\quad h_1 (l)= H (l)-H (l-1),\quad
  H (l) = - \sum_{S_l} p (S_l) \ln p (S_l)
\end{equation}
where the $S_l$ are the allowed symbolic sequences of length $l$, and
the $p (S_l)$ are their probabilities of occurrence in the sequence
associated with a typical chaotic orbit.  In phase space, a good
estimate of $h_1$ can be obtained by a standard Lyapunov exponent
calculation, since it is conjectured that the metric entropy is equal
to the sum of the positive Lyapunov exponents (see,
e.g.,~\cite{Eckmann85a}).

To verify that a partition constructed from a topological analysis of
unstable periodic orbits allows one to compute an accurate estimate of
the metric entropy, we have thus compared estimates obtained by these
two approaches for the regime of the modulated laser system which was
studied in~\cite{PlumecoqP99a}, using the partition $\pa_h^f$ (see
Sec.~\ref{sec:notations}).

First, an estimate of the positive Lyapunov exponent was obtained
in the standard way by integrating the linearized equations of motion
around a numerical solution over a time interval of $10^7$ periods of
modulation. This yielded the value $\lambda_1\approx 0.3730$.

\begin{figure}[htbp]
  \begin{center}
    \leavevmode
    \includegraphics[angle=270,width=12cm]{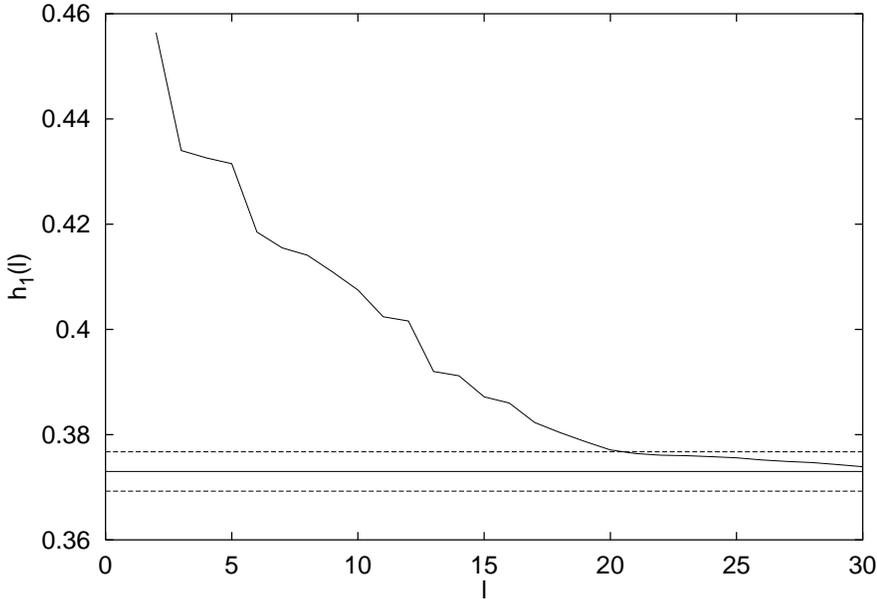}
    \caption{Estimates of the metric entropy using finite sample
    corrections of Eq.~\ref{eq_entropy} vs. string length $l$. The
    horizontal line indicates the value obtained from Lyapunov
    exponent calculations and the dashed lines correspond to $\pm$ 1\%
    deviations from this value.}
    \label{fig:co2-entropy}
  \end{center}
\end{figure}

Then we computed the probabilities $p (S_l)$ for symbol strings of
lengths up to $l_M = 30$ from a sequence of $10^8$ symbols obtained by
encoding a chaotic orbit with the partition $\pa_h^f$. To minimize
systematic errors due to low-probability strings, we applied to
Eq.~(\ref{eq_entropy}) the finite sample corrections derived by
Grassberger~\cite{Grassberger88a}.  The corresponding estimates of
$h_1(l)$ are shown in Fig.~\ref{fig:co2-entropy}.

The best estimate is $h_1\approx h_1(30)= 0.374$ which closely agrees
with $\lambda_1$ as they differ by only $0.3\%$. Furthermore,
estimates within 1\% of the correct value are obtained for $l>20$.  It
is however difficult to give precise error bars. Indeed,
Eq.(\ref{eq_entropy}) systematically overestimates the metric entropy
because it does not take into account the forbidden strings of length
greater than $l_M$. Conversely, the fact that some low-probability
sequences may have not been observed can lead to underestimate the
entropy.

Furthermore, it should be noted that this test is probably not very
sensitive to the quality of the partition: the metric entropy estimate
is maximal for a generating partition and thus should remain almost
constant when the border of the partition is slightly displaced.
Nevertheless, we see it as another check of the validity of our
algorithm.

A perhaps more robust characterization of a chaotic regime is to give
the list of irreducible forbidden words (IFW)~\cite{Badii97a}, which
define the ``grammar'' of this regime. They are the shortest forbidden
symbolic sequences such that every forbidden sequence contains at
least one of them, and can be used to compute topological entropy.  If
we do so for the regime under study, this reveals that it is very
close to the crisis where the attractor collides with a period-$3$
orbit~\cite{Lefranc94a}: the only forbidden sequence up to period 12
is the ``00'' string. For reference, we give in
Table~\ref{tab:co2-forbid} the list of irreducible forbidden words up
to length 17.

\begin{table}[htbp]
    \caption{List of irreducible forbidden words up to length 17 for
      the partition $\pa_h^f$. } 
  \begin{center}
    \label{tab:co2-forbid}
    \leavevmode
    \begin{tabular}[c]{|r|l|}
      \hline
      length & irreducible forbidden words\\
      \hline\hline
      2 & $0^2$ \\
      \hline
      13 & $(01)^2(011)^3$,  $0111(011)^3$\\
      \hline
      15 & $(01)^2(011)^2(01)^2 0$, $0111(011)^2(01)^2 0$, $(011)^5$\\
      \hline
      17 & $(01)^2(011)^20101110$, $0111(011)^20101110$,\\
      & $(011)^4(01)^2 0$,  $1^301(011)^4$,  $1^5(011)^4$\\
      \hline
    \end{tabular}
  \end{center}
\end{table}

Because the list of irreducible forbidden words is an important
characterization of the dynamics, and to allow future comparisons of
our results with these obtained from other methods, we also provide in
Tables \ref{tab:spiral-forbid} and \ref{tab:duffing-forbid} the list
of IFWs for the spiral laser and Duffing attractors studied in
Secs.~\ref{sec:laser_spiral} and \ref{sec:duffing}. In the latter
example, note that symbols ``0'' and ``1'' should be exchanged to
compare the list of IFWs to that of Ref.~\cite{Giovannini91a}.

It can be seen that one of our $7$ period-$6$ IFWs is not listed in
Ref.~\cite{Giovannini91a}: it is ``211100'', which seems to be paired
with ``211101''. We do not presently have an explanation for this
discrepancy, as we showed in Sec.~\ref{sec:homtang_check} that the
partition border we obtain with our approach follows exactly lines of
homoclinic tangencies. We can however note that there is a line of
homoclinic tangencies located slightly below the lower one of
Fig.~\ref{fig:duffing_part}, and which has perhaps been used in
Ref.~\cite{Giovannini91a}.

\begin{table}[htbp]
    \caption{List of irreducible forbidden words up to length 7 for
      the partition of the spiral laser attractor shown in
      Fig.~\ref{fig:spiral_part}, Sec.~\ref{sec:laser_spiral}. } 
  \begin{center}
    \label{tab:spiral-forbid}
    \leavevmode
    \begin{tabular}[c]{|r|l|}
      \hline
      length & irreducible forbidden words\\
      \hline\hline
      2 & $21$, $22$\\
      \hline
      3 & $202$\\
      \hline
      4 & $0000$,$0001$\\
      \hline
      5 & $10201$, $11000$, $11201$, $12000$\\
      \hline
      6 & $1010^3$,  $1020^3$,  $102002$,  $11011$,  $110012$,
      $112002$,  $120011$,  $120012$,  $2010^3$\\
      \hline
      7 & $0^{3}20^{3}$, $1002010$, $1002011$, $1012010$, $1012011$,
      $1100102$, $1200102$,\\
      & $2002010$, $2010011$, $2010012$, $2012010$\\
      \hline
    \end{tabular}
  \end{center}
\end{table}

\begin{table}[htbp]
    \caption{List of irreducible forbidden words up to length 7 for
      the partition the Duffing attractor shown in
      Fig.~\ref{fig:duffing_part}, Sec.~\ref{sec:duffing}. The
      underlined IFW of length $6$ is not listed in
      Ref.~\cite{Giovannini91a}.}   
  \begin{center}
    \label{tab:duffing-forbid}
    \leavevmode
    \begin{tabular}[c]{|r|l|}
      \hline
      length & irreducible forbidden words\\
      \hline\hline
      2 & $12$, $22$\\
      \hline
      4 & $2020$\\
      \hline
      5 & $11111$, $21111$\\
      \hline
      6 & $101111$, $102111$, $111100$, $111101$, \underline{$211100$},
      $211101$, $202102$\\
      \hline
      7 & $1011100$, $1021100$, $2021011$\\
      \hline
    \end{tabular}
  \end{center}
\end{table}

\section{Compact parameterization of partitions}
\label{sec:pract_app}

In the algorithm described in the previous sections, all the detected
periodic points were eventually inserted in the triangulation. While
this ensures that all the available information is utilized, it makes
the description of the final partition complex. Consequently, encoding
is computationally intensive, since one has to find at each iteration
the nearest neighbor of the current state among thousands of reference
points. However, as seen in~\cite{PlumecoqP99a}, only those points
that are in a small neighborhood of the border line are relevant to
specify the partition. In this section, we accordingly show that the
high-resolution partitions we have obtained in the previous sections
can in fact be parameterized by a small number of reference points.

\begin{figure}[htbp]
  \begin{center}
    \leavevmode
    \includegraphics[width=13cm]{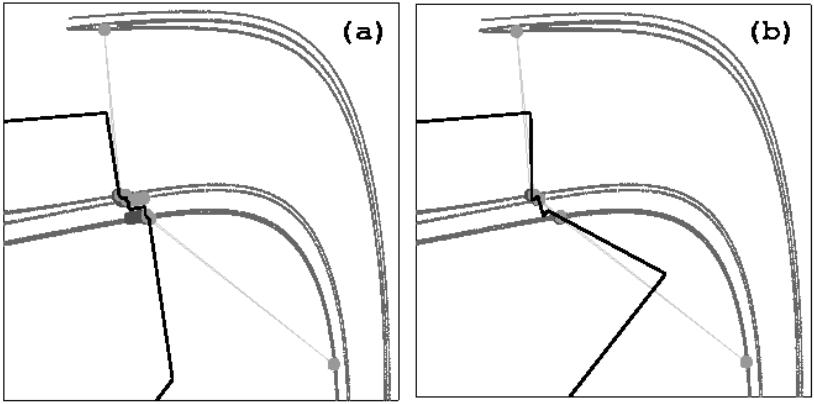}
    \caption{(a) Parameterizing a partition with reference points belonging
      to triangles bearing different symbols and whose mediators are
      located inside the shadow polygon yields a partition that cannot
      be distinguished from the original, yet has a much more compact
      description. (b) The number of reference points can be further
      reduced by keeping only those whose removal would change the
      encoding of one of the detected points.}
    \label{fig:cleaned_part}
  \end{center}
\end{figure}

The border line is built with the mediators of triangles having
vertices with different symbols. Since only the part of the border
line which is located inside the polygon enclosing the strange
attractor is relevant to perform the encoding, the essential
information is actually carried by triangles whose mediators are
located inside this polygon.  Therefore, we can discard reference
points which do not belong to such triangles without modifying the
border of the partition inside the shadow polygon, hence inside the
support of the attractor. In the case of the partition $\pa_a^f$
computed in~\cite{PlumecoqP99a} and displayed in
Fig.~\ref{fig:ref_part}b, this allows one to parameterize the final
partition using only 46 reference points among the 34090 detected
periodic points, as shown in Fig.~\ref{fig:cleaned_part}a. The reason
why the number of reference points remains relatively high is that
triangles that connect different leaves of the attractor, but are
entirely located inside the polygon, cannot be removed by the above
rule whereas they in fact do not carry information about the location
of the border. This limitation is of course due to the naive way in
which the procedure used to decide whether the encoding of a point is
uncertain has been implemented: it was suited to determining the
partition but not to finding its smallest parameterization.

Going back to the core ideas of the algorithm, we can achieve a much
more compact description if we allow the border to be slightly
displaced, but without modifying the encoding of any periodic point.
Consider periodic points located in the Vorono\"\i{} domain $V_p$ of a
given reference point $p$. If the second nearest neighbors of all
these points have the same symbol as $p$, $p$ can be safely removed
without modifying the symbols that the partition assigns to any
periodic point. To simplify the partition, one can therefore search
for such reference points until further removals would change the
encoding of some periodic points.

To achieve a compact parameterization, we have proceeded as follows.
Given a periodic point $O_i^j$ located in region $\region_s(\pa)$, we
define $n_r (O_i^j)$ as being the number of reference points with
symbol $s$ which are closer than the closest reference point with
symbol $s'\neq s$.  Then, the redundancy of each reference point $p$
is estimated by $r (p) = \min_{O\in V_{p}} n_r (O)$. 

We then remove the reference points with the highest $r (p)$ until the
number of removed points is equal to the highest $r(p)$ for the
remaining points, so as to ensure that for each periodic point there
is still a nearest neighbor with the correct symbol. Indeed, a high
value of $r(p)$ indicates that $p$ carries little information, as at
least $r(p)$ reference points, including $p$, can be removed before
the symbols associated to periodic points $O\in V_{p}$ are
modified. This step is repeated until no reference point can be
removed.

We have applied this procedure to the partition shown in
Fig.~\ref{fig:cleaned_part}a, and obtained a simplified partition
(displayed in Fig.~\ref{fig:cleaned_part}b) which is parameterized by
only 8 reference points, yet has the same precision (of the order of
0.1\% of the attractor width) as the original partition $\pa_a^f$
(Fig.~\ref{fig:ref_part}b).

It can be seen on Fig.~\ref{fig:cleaned_part}b that a compromise has
to be found between compactness and robustness to noise. Indeed, with
a small number of reference points, the partition line tends to wiggle
and to pass close to a large number of periodic points, which
increases the probability of encoding errors due to noise.  A
satisfying balance between these two requirements can be tuned for
example by allowing the removal of a reference point $p$ only if
$r(p)>r_{\text{min}}$, where $r_{\text{min}}$ is appropriately chosen.
Our current selection algorithm can probably be much improved.
However, the partitions of Fig.~\ref{fig:cleaned_part} show clearly
that a high resolution can be achieved at a minimal cost. Furthermore,
it should be recalled that in practical applications, such as
chaos-based communication, the system trajectory is usually forced to
stay at a finite distance of the partition border, which should allow
one to simplify the description even more.

To conclude this section, we would like to show with a simple example
that the method we have presented here has modest computational
requirements when we limit ourselves to resolutions relevant to
practical applications, and that it is well suited to perform a
real-time encoding of a chaotic system.

\begin{figure}[htbp]
  \begin{center}
    \leavevmode
    \includegraphics[width=10cm]{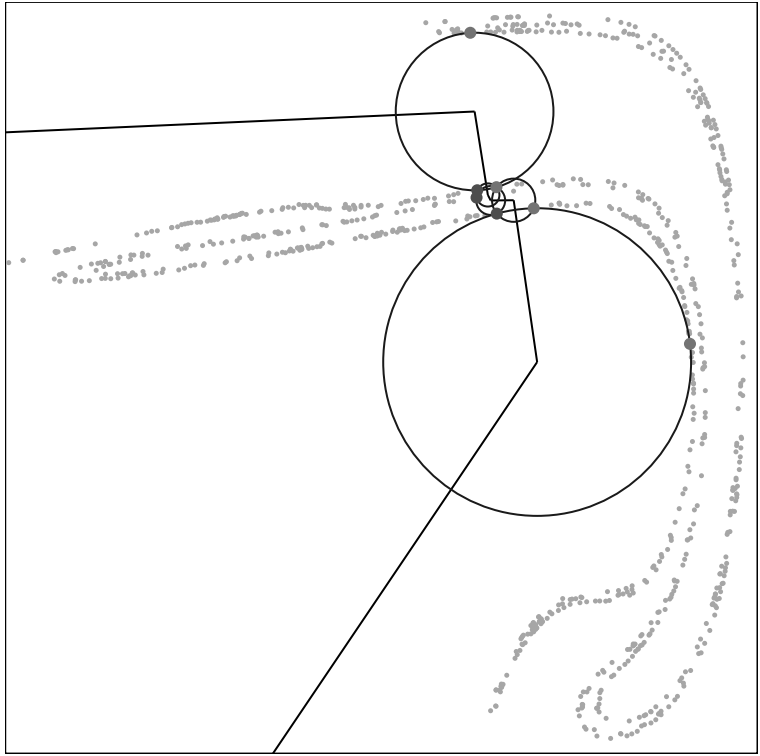}
    \caption{ Partition obtained with a small set of periodic
      orbits (shown in grey), parameterized with only 7 points. The
      circumcenters delimiting the uncertain region are also displayed.}
    \label{fig:exp_part}
  \end{center}
\end{figure}

To this end, we computed a cover of the attractor of
Fig.~\ref{fig:ref_part}b with UPO of period up to $15$, and with a
resolution of 3\%, obtained with 75 periodic orbits. We applied to
this set of UPO the standard algorithm described
in~\cite{PlumecoqP99a}. The resulting partition was then simplified as
described above to obtain the partition of Fig.~\ref{fig:exp_part}
whose description utilizes only $7$ reference points. The
computational costs of the different stages of the algorithm are given
in Table~\ref{tab:comp_cost} and can be seen to be very
small\footnote{It should however be noted that, because of the
  exponential complexity of the problem, the determination of the
  high-resolution partitions presented in the previous sections
  required CPU times of the order of one hundred hours, essentially
  for the detection of UPO.}. Yet, in spite of its much lower
precision, the partition shown in Fig.~\ref{fig:exp_part}b is
characterized by exactly the same irreducible forbidden words as the
high-resolution partition $\pa_h^f$ (see Sec.~\ref{sec:notations}), at
least up to length 19 included. Remarkably, we observed the same
agreement for a similar partition based on 43 orbits of periods up to
$12$ which displayed a much wider gap around the border.

This example clearly shows that experimental systems can be precisely
characterized using the topological approach. It should be noted that
for regimes that have a higher topological entropy, such as these
found after the period-$3T$ crisis as in Ref.~\cite{Lefranc94a}, a
higher precision can be obtained using lower-period orbits. Also, the
partition shown in Ref.~\cite{Lefranc94a} shows that even if most of
the periodic orbits of period larger than 10 are very difficult to
detect, this is in general not the case for those located just near
the border line, as they are weakly unstable.

\begin{table}[htbp]
  \caption{Computational cost of the different stages of this
    algorithm for the partition of Fig.~\ref{fig:exp_part}. CPU times
    are given for a Pentium 133 computer running under the Linux
    operating system.} 
  \label{tab:comp_cost}
  \begin{center}
    \leavevmode
    \begin{tabular}[c]{|lr|}
      \hline
      Task & CPU time (seconds)\\
      \hline\hline
      UPO detection & 287.5\\
      Computation of topological invariants & 47.9\\
      Template determination & 0.9 \\
      Determination of topological names & 2.0 \\
      Construction of the partition & 5.4\\
      \hline
    \end{tabular}
  \end{center}
\end{table}

\section{Conclusion}
\label{sec:discussion_conclusion}

In this work, we have carried out precise tests of the validity of the
algorithm presented in~\cite{PlumecoqP99a}.  A first indication that
the resulting encodings faithfully describe the dynamics is that
partitions determined from different initial configurations are
equivalent: they encode a given trajectory in the same way, up to a
shift of the symbolic sequences (Sec.~\ref{sec:choice}).  Furthermore,
we have also directly verified that these partitions closely meet the
criteria for being a generating partition by (i) observing that
symbolic sequences of increasing length were associated to regions of
decreasing diameters (Sec.~\ref{sec:generating_test}), and (ii) by
obtaining good estimates for the metric entropy from a long chaotic
sequence (Sec.~\ref{sec:entropy}).

However, the most convincing evidence of the relevance of our method
is provided by the perfect agreement between our results and those
obtained by the classical method based on homoclinic tangencies.
Indeed, we have found that important lines of homoclinic tangencies
are entirely located inside the triangles enclosing the border of the
partition (Sec.~\ref{sec:homtang_check}). This does not only mean that
the partitions we obtain are very close to what is conjectured to be
the correct solution, but also that our method naturally provides
reliable error bounds, which are determined from the circumcircles of
the triangles enclosing the partition border.

Even if the ultimate precision we can reach is probably lower than
with homoclinic tangencies, the topological approach to constructing
symbolic encodings has several distinctive advantages.

First, the problem of how to choose the tangency lines that will
define a good partition is naturally solved. This was illustrated by
the study of the Duffing attractor (Sec.~\ref{sec:duffing}). In this
example, our algorithm naturally selects the lines of homoclinic
tangencies that were found in Ref.~\cite{Giovannini91a} to yield the
correct partition although some of them do not correspond to lines of
primary homoclinic tangencies. That the solutions so obtained are the
most natural ones is confirmed by the results of Sec.~\ref{sec:choice}
where it was found that the different solutions were images or
preimages of each other.

Second, the determination of the underlying template automatically
indicates the correct number of symbols on which the dynamics is
based.  This is because this number is nothing but the number of
branches of the template. This is also a key property since we have
seen that for a given system, the number of symbols may depend on the
parameter values as is the case, e.g., for the modulated laser system
(compare the partitions shown in Figs.~\ref{fig:ref_part}b and
\ref{fig:spiral_part}).

Last but not least, the topological approach should be extremely
robust to noise. Indeed, it makes use of trajectories located
in the whole phase space, and only gradually converges to the lines of
homoclinic tangencies, where it is known that a dramatic noise
amplification takes place~\cite{Jaeger97a}. The latter phenomenon is
likely to adversely affect the determination of the directions of the
invariant manifolds in the neighborhood of the tangencies from the
time series, especially as the most expanding direction is orthogonal
to the unstable manifold~\cite{Jaeger97a} there.

In contrast with this, the very fact that at homoclinic tangencies
noise perturbs trajectories orthogonally to the unstable manifold
implies that the locations of periodic points as estimated from close
returns will be in first approximation displaced in a direction that
is parallel to the border of the partition. The latter should thus be
minimally affected, unless noise is so strong as to make the periodic
point appear to be in different leaf of the attractor than the one it
belongs to. Furthermore, it should be noted that the determination of
topological invariants is relatively insensitive to noise level as the
knot type of an orbit generally still can be exactly determined from
an noisy, approximate trajectory shadowing this orbit.

All these properties clearly show that the topological approach should
be a method of choice to construct generating partitions from
experimental time series. As a matter of fact, we recall that a
preliminary version of our algorithm has already been successfully
applied to obtain with a very good accuracy a generating partition for
a chaotic attractor of a modulated laser~\cite{Lefranc94a}. The
relevance of template analysis to study the symbolic dynamics of
experimental systems is further confirmed by the observation in
Secs.~\ref{sec:generating_test} and \ref{sec:pract_app} that a
precision of only a few percent is sufficient to characterize very
precisely the grammar of a chaotic regime.

We believe that our algorithm, and the way in which partitions are
parameterized, could also prove useful in practical applications, and
especially for transmitting information over chaotic signals, as
proposed by Hayes, Grebogi and Ott~\cite{Hayes93a,Hayes94a}. In this
approach, the message to be transmitted is first encoded into a
sequence of symbols that is compatible with the symbolic dynamics of
the transmitting device. Then, tiny perturbations are applied to this
device so that among all possible chaotic trajectories, it follows the
one whose associated symbolic sequence is the given sequence. Indeed,
we showed in Sec.~\ref{sec:pract_app} that once the partition has been
determined, it is possible to simplify very much its description, so
that encoding a signal merely amounts to finding among a few reference
points which is closest to the point in phase space representing the
current state of the system. This computation could easily be
implemented in a digital signal processor, to achieve real-time
encoding at the receiving side of the transmission line and recover
thereby the digital message.

The fact that the images of a given partition can be easily and
exactly computed (reference points are periodic) would also be
beneficial at the sending side because it makes it easier to predict
the next symbols that the device would emit if it was free-running,
and thus to compute the tiny perturbation needed to synchronize the
symbolic stream with the prescribed sequence. Finally, we note that
our algorithm naturally selects a region where the encoding is to be
considered uncertain, and which the transmitting device should not
visit in order to avoid any ambiguity at the receiving side.

In conclusion, we believe that the present work shows that topological
analysis, in the present form or in future extensions, is a promising
tool to master the symbolic dynamics of experimental chaotic systems,
especially given its robustness to noise and its ability to select the
appropriate tangency lines and the correct number of symbols.

\begin{ack}
  It is a pleasure to thank our colleagues Guillaume Boulant, Serge
  Bielawski, Dominique Derozier, and Robert Gilmore for stimulating
  discussions.  The Laboratoire de Physique des Lasers, Atomes,
  Mol\'ecules is Unit\'e de Recherche Mixte du CNRS. The Centre
  d'\'Etudes et Recherches Lasers et Applications is supported by the
  Minist\`ere charg\'e de la Recherche, the R\'egion Nord-Pas de
  Calais and the Fonds Europ\'een de D\'eveloppement \'Economique des
  R\'egions.
\end{ack}

\end{document}